\documentclass[11pt]{article}%
\usepackage{amsmath}
\usepackage{moriond,epsfig}
\usepackage{graphicx}
\usepackage{amsfonts}
\usepackage{amssymb}%
\def\be{\begin{equation}}
\def\ee{\end{equation}}
\def\bea{\begin{eqnarray}}
\def\eea{\end{eqnarray}}

\begin{document}
%
\vspace*{4cm}
\title{TRACKING DOWN PENGUINS AT THE POLES}
\author{ J.-M. GERARD,${}^{1}~$ C. SMITH,${}^{2}~$ S. TRINE ${}^{3,a}%
~$\footnotetext[1]{Speaker} \\[4pt]}
\address{
${}^{1}~$Institut de Physique Th\'{e}orique, Universit\'{e}
catholique de Louvain,\\
Chemin du Cyclotron 2, B-1348 Louvain-la-Neuve, Belgium\\[3pt]
${}^{2}%
~$Institut f\"ur Theoretische Physik, Universit\"at Bern, CH-3012 Bern, Switzerland\\[3pt]
${}^{3}%
~$Institut f\"ur Theoretische Teilchenphysik, Universit\"at Karlsruhe, D-76128 Karlsruhe, Germany}
\maketitle\abstracts{QCD penguins are responsible for about 2/3 of the $\Delta
I=1/2$ rule
in $K\rightarrow\pi\pi
$ decays, as inferred from a combined analysis of $K\rightarrow\pi
\pi$ and $K_L\rightarrow\gamma\gamma$.
Further tests based on the decays
$K_S\rightarrow\pi^0\gamma\gamma$ and $K^{+}\rightarrow\pi^{+}\gamma
\gamma$ are proposed.
New insights into the treatment of $\pi^0$, $\eta$, $\eta
'$ pole amplitudes are also reported.}%

\section{Introduction}

Recently, a systematic analysis of $\eta_{0}$ pole contributions to radiative
$K$ decays was performed in the context of large $N_{c}$ ChPT, in order to
better understand the role of gluonic penguin operators in $K\rightarrow\pi
\pi$ transitions\thinspace\cite{GerardST05}. In this note, we emphasize some
aspects of this study, in view of the forthcoming new experimental information
on $K^{+}\rightarrow\pi^{+}\gamma\gamma$ by the NA48 Collaboration\thinspace
\cite{NA4806}. A number of issues, like the correspondence between the $SU(3)$ and
$U(3)$ chiral expansions, the impact of our analysis for $K_{L}\rightarrow
\gamma\gamma^{\ast}$, $K_{L}\rightarrow\pi^{0}\pi^{0}\gamma\gamma$ and
$K_{L}\rightarrow\pi^{+}\pi^{-}\gamma$, or the fate of the weak mass operator,
are left to the original paper.

\section{General framework}

The effective weak Hamiltonian relevant to describe (CP-conserving) hadronic
$K$ decays reads:%
\begin{equation}
\mathcal{H}_{eff}^{\Delta S=1}\left(  \mu<m_{c}\right)  \simeq\frac{G_{F}%
}{\sqrt{2}}V_{ud}V_{us}^{\ast}\left[  z_{1}\left(  \mu\right)  Q_{1}\left(
\mu\right)  +z_{2}\left(  \mu\right)  Q_{2}\left(  \mu\right)  +z_{6}\left(
\mu\right)  Q_{6}\left(  \mu\right)  \right]  ,\label{WeakEffHam}%
\end{equation}
with the familiar current-current operators%
\begin{equation}%
\begin{tabular}
[c]{cc}%
$Q_{1}=4\left(  \bar{s}_{L}\gamma_{\alpha}d_{L}\right)  \left(  \bar{u}%
_{L}\gamma^{\alpha}u_{L}\right)  ,$ & $Q_{2}=4\left(  \bar{s}_{L}%
\gamma_{\alpha}u_{L}\right)  \left(  \bar{u}_{L}\gamma^{\alpha}d_{L}\right)
,$%
\end{tabular}
\end{equation}
and the density-density dominant penguin operator%
\begin{equation}%
\begin{tabular}
[c]{c}%
$Q_{6}=-8\left(  \bar{s}_{L}q_{R}\right)  \left(  \bar{q}_{R}d_{L}\right)  .$%
\end{tabular}
\end{equation}
In our notations, $q_{L}^{R}\equiv\frac{1}{2}(1\pm\gamma_{5})q$ and the light
flavours $q=u,d,s$ are summed over. The effective coupling constants
$z_{i}\left(  \mu\right)  $ contain QCD effects above the renormalization
scale $\mu$, kept high enough to allow the use of perturbation theory. In
order to investigate the effects of long-distance strong interactions, we will make use of
ChPT (Chiral Perturbation Theory) techniques.

ChPT relies on the $SU(3)_{L}\times SU(3)_{R}$ symmetry of the QCD Lagrangian
in the massless limit to build a dual representation, in terms of meson
fields. If one formally considers the number of QCD colours $N_{c}$ as large,
$SU(3)$ can be extended to $U(3)$ and the spontaneous symmetry breaking
$U(3)_{L}\times U(3)_{R}\rightarrow U(3)_{V}$ gives rise to a nonet $\Pi$ of
pseudoscalar Goldstone bosons, which are written $U\equiv\exp (i\sqrt{2}
\Pi / F)$ in the standard parametrization. This extension to $U(3)$ will prove
crucial afterwards. The corresponding leading nonlinear Lagrangian reads
\begin{equation}
\mathcal{L}_{S}^{(p^{2},\infty)+(p^{0},1/N_{c})}=\dfrac{F^{2}}{4}%
\langle\partial_{\mu}U\partial^{\mu}U^{\dagger}\rangle+\dfrac{F^{2}}{4}%
\langle\chi U^{\dagger}+U\chi^{\dagger}\rangle+\dfrac{F^{2}}{16N_{c}}m_{0}%
^{2}\langle\ln U-\ln U^{\dagger}\rangle^{2}\label{LagrNL}%
\end{equation}
where $\left\langle {}\right\rangle $ denotes a trace over flavours, the
external source $\chi$ is frozen at $\chi=rM$ with $M=diag(m_{u},m_{d},m_{s})$
to account for meson masses, $F$ is identified with the pion decay constant
$F_{\pi}=92.4$ MeV at this order and $m_{0}$ represents the anomalous part of
the $\eta_{0}$ mass. Note that the leading $SU(3)$ chiral Lagrangian is
recovered in the limit $m_{0}\rightarrow\infty$, when the $\eta_{0}$ decouples.

The meson realization of Eq.(\ref{WeakEffHam}) can be obtained from the chiral
representations of the corresponding quark currents and densities, i.e.,
preserving the colour and flavour structures:
\begin{equation}
\mathcal{H}_{eff,\mathcal{O}(p^{2})}^{\Delta S=1}\left(  \mu\sim m_{\pi
,K}\right)  \simeq\frac{G_{F}}{\sqrt{2}}V_{ud}V_{us}^{\ast}\left[  x_{1}%
\hat{Q}_{1}+x_{2}\hat{Q}_{2}+x_{6}\hat{Q}_{6}\right]  ,\label{WeakEffHamLD}%
\end{equation}
with%
\begin{equation}%
\begin{tabular}
[c]{ccc}%
$\hat{Q}_{1}=4\left(  L_{\mu}\right)  _{23}\left(  L^{\mu}\right)  _{11},$ &
$\hat{Q}_{2}=4\left(  L_{\mu}\right)  _{13}\left(  L^{\mu}\right)  _{21},$ &
$\hat{Q}_{6}=4\left(  L_{\mu}L^{\mu}\right)  _{23},$%
\end{tabular}
\end{equation}
and the left-handed currents $\left(  L_{\mu}\right)  ^{lk}\equiv i\frac
{F^{2}}{2}\left(  \partial_{\mu}UU^{\dagger}\right)  ^{lk}.$ The weak
coefficients $x_{i}$ are not fixed by symmetry arguments, and contain both
short-distance and long-distance strong interaction effects. The latter are
known to be important in explaining the $\Delta I=1/2$ rule observed in
$K\rightarrow\pi\pi$ decays\thinspace\cite{BardeenBG}. Still, the genuine
mechanism responsible for the $\Delta I=1/2$ enhancement, i.e., the relative
strength of the penguin and current-current operators, has not been completely
settled yet. In this work, we propose a phenomenological extraction of the
$x_{i}$ parameters, and thus of the penguin fraction $\mathcal{F}_{P}%
=3x_{6}/(-x_{1}+2x_{2}+3x_{6}).$

To reach this goal, it is clear that one has to go beyond the standard $SU(3)$
ChPT which only contains two independent weak operators ($Q_{8}$ and $Q_{27}$)
such that current-current and penguin operators cannot be disentangled. On the
other hand, in $U(3)$, the presence of $\eta_{0}$ as a dynamical degree of
freedom allows for an extra $\mathcal{O}(p^{2})$ weak operator%
\begin{equation}
Q_{8}^{s}=4\left(  L_{\mu}\right)  _{23}\left\langle L^{\mu}\right\rangle
\sim\left(  L_{\mu}\right)  _{23}\,\partial^{\mu}\eta_{0},\label{Q8s}%
\end{equation}
which, together with the straightforward extensions of $Q_{8}$ and $Q_{27}$ to
$U(3)$%
\begin{equation}%
\begin{tabular}
[c]{cc}%
$Q_{8}=4\left(  L_{\mu}L^{\mu}\right)  _{23},$ & $Q_{27}=4\left[  \left(
L_{\mu}\right)  _{23}\left(  L^{\mu}\right)  _{11}+\frac{2}{3}\left(  L_{\mu
}\right)  _{13}\left(  L^{\mu}\right)  _{21}-\frac{1}{3}\left(  L_{\mu
}\right)  _{23}\left\langle L^{\mu}\right\rangle \right]  ,$%
\end{tabular}
\end{equation}
permits now to write the effective Hamiltonian in a way equivalent to
Eq.(\ref{WeakEffHamLD}):
\begin{equation}
\mathcal{H}_{eff,\mathcal{O}(p^{2})}^{\Delta S=1}\left(  \mu\sim m_{\pi
,K}\right)  \simeq G_{8}Q_{8}+G_{27}Q_{27}+G_{8}^{s}Q_{8}^{s}.
\end{equation}
Explicitly, this change of basis reads ($G_{W}\equiv G_{F}V_{ud}V_{us}^{\ast
}/\sqrt{2}$):%
\begin{equation}%
\begin{tabular}
[c]{ccc}%
$G_{8}/G_{W}=-\frac{2}{5}x_{1}+\frac{3}{5}x_{2}+x_{6},$ & $G_{8}^{s}%
/G_{W}=\frac{3}{5}x_{1}-\frac{2}{5}x_{2},$ & $G_{27}/G_{W}=\frac{3}{5}\left(
x_{1}+x_{2}\right)  .$%
\end{tabular}
\end{equation}
$G_{8}$ and $G_{27}$ are still extracted from $K\rightarrow\pi\pi$. The
knowledge of $G_{8}^{s}$ would thus give access to the $x_{i}$ parameters, and
consequently to $\mathcal{F}_{P}$. Because of Eq.(\ref{Q8s}), natural
candidates for its extraction are anomaly-driven radiative $K$ decays, that
receive a $\eta_{0}$ pole contribution.

\section{Penguin fraction in $K\rightarrow\pi\pi$ vs $\eta_{0}$ effects in
$K_{L}\rightarrow\gamma\gamma$}

Due to the well-known pole cancellations at work in $K_{L}\rightarrow
\gamma\gamma$, we propose a two-step analysis for this mode:\medskip

\emph{Step 1}: work with the \emph{theoretical} masses $m_{\pi}$, $m_{\eta}$,
$m_{\eta^{\prime}}$, i.e., consistently at a given order in ChPT, in order to
identify the vanishing pole contributions (Fig.1a). It turns out that $Q_{8}$
does not contribute at $\mathcal{O}(p^{4})$, just like in $SU(3)$ ChPT. The
leading contribution, of $\mathcal{O}(p^{4})$, comes from the $\overline{u}u$
intermediate state generated by $\hat{Q}_{1}$ (Fig.1b), and is thus proportional to $G_Wx_1=G_{8}^{s}+2G_{27}/3$,
i.e., the nonet-symmetry breaking couplings.%
\begin{figure}
[t]
\begin{center}
\includegraphics[
height=2.8692cm,
width=10.4948cm
]%
{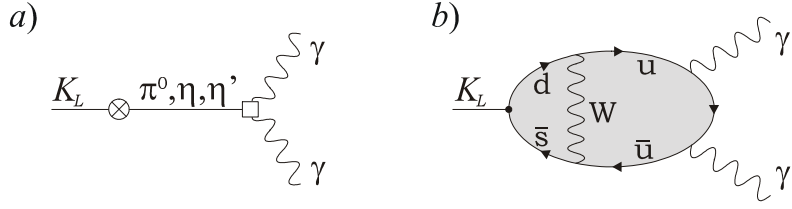}%
\caption{a) Pole diagrams for $K_{L}\rightarrow\gamma\gamma$. b) Dominant
long-distance $\bar{u}u$ contribution.}%
\end{center}
\end{figure}
\medskip

\emph{Step 2}: freeze the $\pi^{0}$, $\eta$, $\eta^{\prime}$ poles at the
\emph{physical} values $M_{\pi}$, $M_{\eta}$, $M_{\eta^{\prime}}$ to ensure
correct analytical properties for the remaining contributions ($\hat{Q}_{1}$)
only. This is done through the following prescription for the $\eta$%
-$\eta^{\prime}$ propagator:%
\begin{equation}
iP_{phys}\left(  q^{2}\right)  _{\eta_{8}\eta_{0}}^{-1}=\left(
\begin{array}
[c]{cc}%
\cos\theta_{P} & \sin\theta_{P}\\
-\sin\theta_{P} & \cos\theta_{P}%
\end{array}
\right)  \left(
\begin{array}
[c]{cc}%
q^{2}-M_{\eta}^{2} & 0\\
0 & q^{2}-M_{\eta^{\prime}}^{2}%
\end{array}
\right)  \left(
\begin{array}
[c]{cc}%
\cos\theta_{P} & -\sin\theta_{P}\\
\sin\theta_{P} & \cos\theta_{P}%
\end{array}
\right)  \,, \label{PMP3}%
\end{equation}
where the parametrisation in terms of one mixing angle is allowed as we work
at lowest order in the chiral expansion, cf. Eq.(\ref{LagrNL}). A discussion
of two-angle pole formulas may be found in our original paper \cite{GerardST05}%
.\medskip

The resulting pole amplitude ($c_{\theta}\equiv\cos\theta_{P}$, $s_{\theta
}\equiv\sin\theta_{P}$),
\begin{align}
&  \mathcal{A}^{\mu\nu}\left(  K_{L}\rightarrow\gamma\gamma\right)
=\frac{2F\alpha}{\pi}\left(  G_{8}^{s}+\dfrac{2}{3}G_{27}\right)  M_{K}%
^{2}i\varepsilon^{\mu\nu\rho\sigma}k_{1\rho}k_{2\sigma}\nonumber\\
&  \;\;\;\times\left(  \frac{1}{M_{K}^{2}-M_{\pi}^{2}}+\frac{(c_{\theta
}-2\sqrt{2}s_{\theta})(c_{\theta}-\sqrt{2}s_{\theta})}{3(M_{K}^{2}-M_{\eta
}^{2})}+\frac{(s_{\theta}+2\sqrt{2}c_{\theta})(s_{\theta}+\sqrt{2}c_{\theta}%
)}{3(M_{K}^{2}-M_{\eta^{\prime}}^{2})}\right)  \;, \label{KL12}%
\end{align}
turns out to be dominated by the $\eta$:%
\begin{equation}
\mathcal{A}^{\mu\nu}\left(  K_{L}\rightarrow\gamma\gamma\right)  =\left(
G_{8}^{s}+\dfrac{2}{3}G_{27}\right)  \left[  \left(  0.46\right)  _{\pi
}-\left(  1.83\pm0.30\right)  _{\eta}-\left(  0.12\pm0.02\right)
_{\eta^{\prime}}\right]  i\varepsilon^{\mu\nu\rho\sigma}k_{1\rho}k_{2\sigma
}\;, \label{KL13}%
\end{equation}
and is quite stable with respect to the $\eta_{8}$-$\eta_{0}$ mixing angle
$\theta_{P}$, allowed to vary in the large range $\left[  -25%
{{}^\circ}%
,-15%
{{}^\circ}%
\right]  $ to get a hold on the typical size of NLO effects. From the
experimental $K_{L}\rightarrow\gamma\gamma$ branching ratio\thinspace
\cite{KLgg}, we obtain $\left(  G_{8}^{s}/G_{8}\right)  _{ph}\simeq\pm1/3,$ in
agreement with the QCD-inspired value\thinspace\cite{GerardST05}%
\begin{equation}
\left(  G_{8}^{s}/G_{8}\right)  _{th}=-0.38\pm0.12, \label{QCDinspG8s}%
\end{equation}
leading to $\left(  \mathcal{F}_{P}\right)  _{th}\simeq60\%$.

\section{$K_{S}\rightarrow\pi^{0}\gamma\gamma$ - the simplest probe}

The simplest mode to test $\left(  G_{8}^{s}/G_{8}\right)  _{th+ph}\simeq-1/3$
is $K_{S}\rightarrow\pi^{0}\gamma\gamma$. Indeed, at leading order in the
chiral expansion, i.e., $\mathcal{O}(p^{4})$, it proceeds entirely through
pole diagrams (Fig.2a). It receives contributions from $\hat{Q}_{1}$ and
$\hat{Q}_{6}$, but not $\hat{Q}_{2}$, or correlated contributions from
$Q_{8}^{s}$, $Q_{27}$ and $Q_{8}$ in the natural $U(3)$ basis. The latter
dominates the decay via the pion pole. When $\eta_{0}$ effects are integrated
out, the standard $SU(3)$ result\thinspace\cite{EckerPR87} is recovered:
\begin{equation}
\mathcal{B}\left(  K_{S}\rightarrow\pi^{0}\gamma\gamma\right)  _{m_{\gamma
\gamma}>220\text{ MeV}}^{SU\left(  3\right)  ,\mathcal{O(}p^{4})}%
=3.8\times10^{-8}\;.\label{Ks9}%
\end{equation}
However, the contribution of the $\eta_{0}$ meson, despite non-leading, can
significantly enhance the branching fraction (Fig.2b).%
\begin{figure}
[t]
\begin{center}
\includegraphics[
height=4.2005cm,
width=12.9252cm
]%
{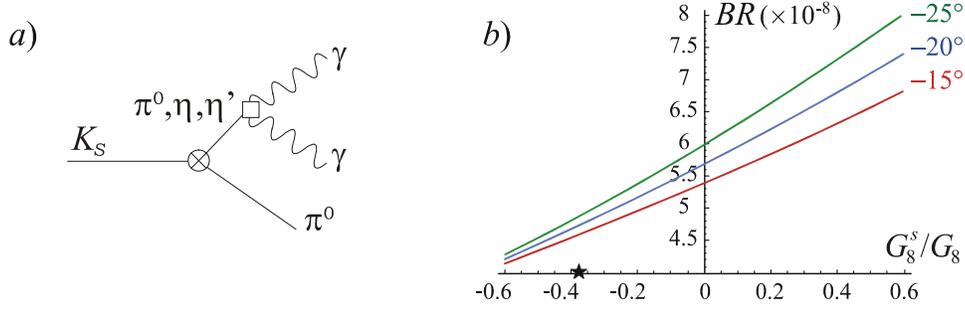}%
\caption{a) Pole diagrams for $K_{S}\rightarrow\pi^{0}\gamma\gamma$. b)
$\mathcal{B}\left(  K_{S}\rightarrow\pi^{0}\gamma\gamma\right)  $ as a
function of $G_{8}^{s}/G_{8}$ for $\theta_{P}=-15{{}^\circ},-20{{}^\circ}%
,-25{{}^\circ}$. The star refers to the theoretical value given in
Eq.(\ref{QCDinspG8s}).}%
\end{center}
\end{figure}
For our preferred value (\ref{QCDinspG8s}) and $\theta_{P}\in\left[  -25%
{{}^\circ}%
,-15%
{{}^\circ}%
\right]  $, we obtain:%
\begin{equation}
\mathcal{B}\left(  K_{S}\rightarrow\pi^{0}\gamma\gamma\right)  _{m_{\gamma
\gamma}>220\text{ MeV}}^{U\left(  3\right)  ,\mathcal{O}(p^{4})}=\left(
4.8\pm0.5\right)  \times10^{-8}\;,\label{Ks11}%
\end{equation}
where the theoretical error only reflects the ranges assigned to $G_{8}%
^{s}/G_{8}$ and $\theta_{P}$. The current experimental value is\thinspace
\cite{NA48KS}:%
\begin{equation}
\mathcal{B}\left(  K_{S}\rightarrow\pi^{0}\gamma\gamma\right)  _{m_{\gamma
\gamma}>220\text{ MeV}}^{\exp}=\left(  4.9\pm1.8\right)  \times10^{-8}%
\ .\label{Ks1}%
\end{equation}
Note that a more precise measurement could already fix the sign of $G_{8}%
^{s}/G_{8}$.

\section{$K^{+}\rightarrow\pi^{+}\gamma\gamma$ - a promising probe}

The case of $K^{+}\rightarrow\pi^{+}\gamma\gamma$ is slightly more involved as
it proceeds through both loop and pole diagrams at leading order in the chiral
expansion, i.e., again, $\mathcal{O}(p^{4})$. Still, these two types of
contributions correspond to photons in different $CP$ eigenstates, and do not
interfere in the rate. The usual $SU(3)$ analysis, including unitarity
corrections, can thus be applied to the loops while the poles (Fig.3a),
sensitive to $\eta_{0}$ effects, are better treated within the $U(3)$ framework. \nopagebreak

Unlike for $K_{S}\rightarrow\pi^{0}\gamma\gamma$, the pion pole contribution
from $Q_{8}$ plays a minor role here as $K^{+}\rightarrow\pi^{+}\pi^{0}$ is
purely $\Delta I=3/2$ when on-shell. The pole amplitude is thus quite
sensitive to $Q_{8}^{s}$ and $Q_{27}$. Already at the $SU(3)$ level, when
$\eta_{0}$ effects are discarded, one can see that the $27$ operator actually
accounts for about half of the pole-induced branching fraction:%
\begin{equation}
\mathcal{B}\left(  K^{+}\rightarrow\pi^{+}\gamma\gamma\right)  _{m_{\gamma
\gamma}>220\text{ MeV}}^{\mathrm{P},SU(3),\mathcal{O}(p^{4})}=1.17\times
10^{-7}\;,\label{Kp11}%
\end{equation}
instead of $0.51\times10^{-7}$ without $Q_{27}$\thinspace\cite{EckerPR88}. The
contribution of the $\eta_{0}$ meson may substantially suppress or enhance
this value, depending on $G_{8}^{s}/G_{8}$ (Fig.3b).%

\begin{figure}
[t]
\begin{center}
\includegraphics[
height=8.3967cm,
width=15.1873cm
]%
{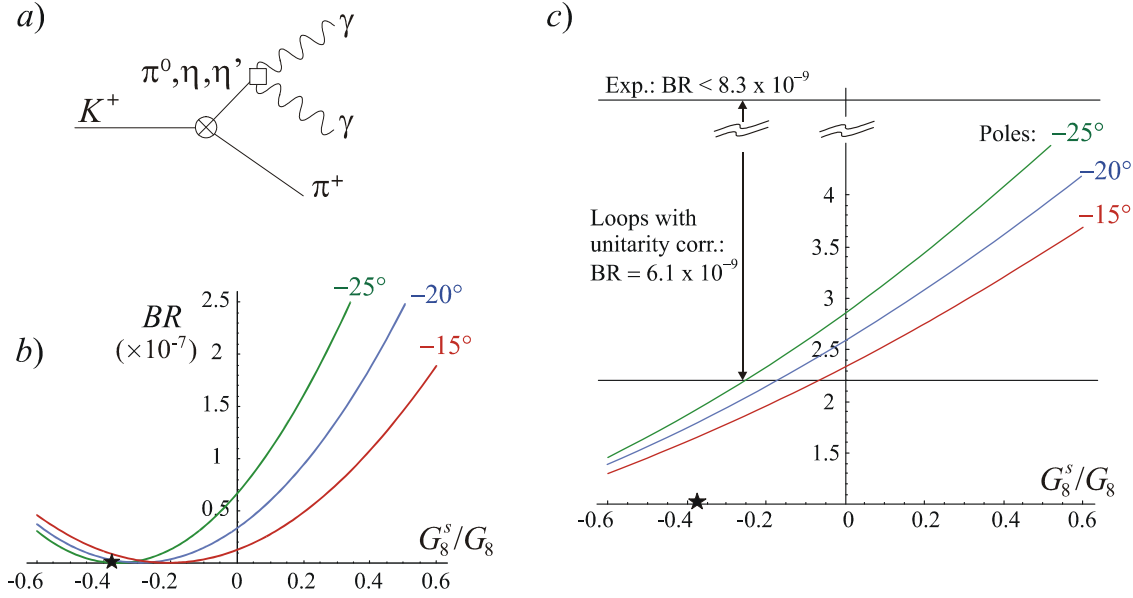}%
\caption{a) Pole diagrams for $K^{+}\rightarrow\pi^{+}\gamma\gamma$. b)
$\mathcal{B}\left(  K^{+}\rightarrow\pi^{+}\gamma\gamma\right)  ^{poles}$ as a
function of $G_{8}^{s}/G_{8}$ for $\theta_{P}=-15{{}^\circ},-20{{}^\circ}%
,-25{{}^\circ}$. c) $\mathcal{B}\left( K^{+}\rightarrow\pi^{+}\gamma
\gamma\right) ^{poles}$ for $m_{\gamma\gamma}<108$ MeV, $\times\,10^{9}$.
Assuming non-negligible loop contributions$^8$, the
recent upper bound$^{10}$ hints towards negative values for
$G_{8}^{s}/G_{8}$. The stars refer to Eq.(\ref{QCDinspG8s}).}%
\end{center}
\end{figure}
In particular, for $G_{8}^{s}/G_{8}=-0.38\pm0.12$ and $\theta_{P}\in\left[
-25%
{{}^\circ}%
,-15%
{{}^\circ}%
\right]  $, poles can be safely neglected with respect to loops:%
\begin{equation}
\mathcal{B}\left(  K^{+}\rightarrow\pi^{+}\gamma\gamma\right)  _{m_{\gamma
\gamma}>220\text{ MeV}}^{\mathrm{P},U(3),\mathcal{O}(p^{4})}\lesssim
0.3\times10^{-7}\;,\label{Kp13}%
\end{equation}
while, for $G_{8}^{s}/G_{8}>0$, they could increase the total rate by more
than 20\%. In that case, they should be taken into account in the extraction
of the $\mathcal{O}(p^{4})$ combination of counterterms $\hat{c}$ to reach
consistency between the total and differential rates\thinspace
\cite{EckerPR88,DAmbrosioP,E787}.

Finally, restricting the analysis to the low energy end of the $\gamma\gamma$
spectrum, negative values of $G_{8}^{s}/G_{8}$ are already favoured (cf. Fig3c).

\section{Implication for $\Delta M_{K}$}

Pole diagrams also play a central role in the long distance contribution to
the $K_{L}$-$K_{S}$ mass difference $\Delta M_{K}$ (Fig.4a). The situation
here is quite similar to the one of $K_{L}\rightarrow\gamma\gamma$, in that
the contribution of $Q_{8}$ vanishes both in $SU(3)$ and $U(3)$ ChPT at
$\mathcal{O}(p^{4})$, the leading effect being driven by the $\hat{Q}_{1}$
operator, i.e., a $u\overline{u}$ pair (Fig.4b). The resulting pole formula
was worked out in our original paper\thinspace\cite{GerardST05}. Its
contribution to $\Delta M_{K}$ is summarized in Fig.4c. For the preferred
value Eq.(\ref{QCDinspG8s}), the negative contribution of poles partially
cancels the positive contribution of $\pi\pi$ loops, leaving to short-distance
effects\thinspace\cite{UliePaper} the task of reproducing the bulk of the
observed mass difference.%
\begin{figure}
[t]
\begin{center}
\includegraphics[
height=4.5306cm,
width=13.9072cm
]%
{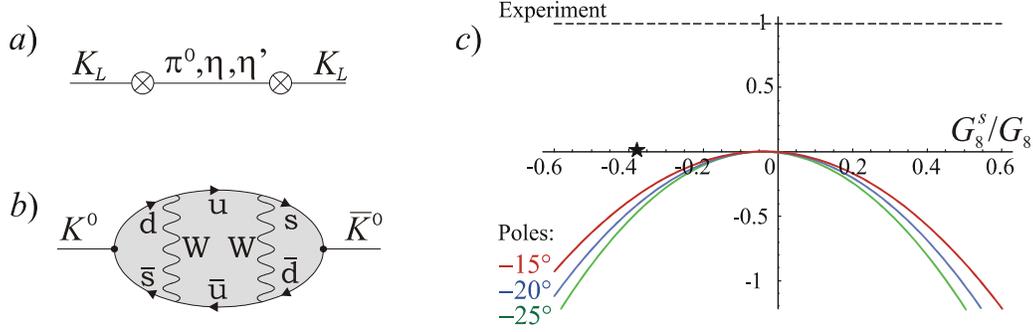}%
\caption{a) Pole diagrams for $\Delta M_{K}^{LD}$. b) Long-distance $\bar{u}u$
contribution. c) Fraction of pole contribution to $\Delta M_{K}^{\exp}$ as a
function of $G_{8}^{s}/G_{8}$ for $\theta_{P}=-15{{}^\circ}$, $-20{{}^\circ}$,
$-25{{}^\circ}$. The star refers to Eq.(\ref{QCDinspG8s}).}%
\end{center}
\end{figure}

\section{Conclusion}

The $\Delta S=1$ effective operator $Q_{8}^{s}$, which describes pure
$\eta_{0}$ effects, holds the key to a phenomenological extraction of the
penguin fraction in $K\rightarrow\pi\pi$ amplitudes via the change of chiral
basis $(\hat{Q}_{1},\hat{Q}_{2},\hat{Q}_{6})\leftrightarrow(Q_{8},Q_{27}%
,Q_{8}^{s})$.

From $\mathcal{B}(K_{L}\rightarrow\gamma\gamma)$, we found $G_{8}^{s}%
/G_{8}\simeq-1/3$, which corresponds to a rather smooth non-perturbative
current-current operator evolution and a penguin contribution to the $\Delta
I=1/2$ rule around $2/3$ at the hadronic scale. Better measurements of the
decays $K_{S}\rightarrow\pi^{0}\gamma\gamma$ and $K^{+}\rightarrow\pi
^{+}\gamma\gamma$ would provide important tests of this picture.

The recourse to (broken) $U(3)$ chiral symmetry also allowed us to identify
correctly the leading contribution to $K_{L}\rightarrow\gamma\gamma$, namely
the transition $K_{L}\rightarrow u\overline{u}$ generated by $\hat{Q}_{1}$.
This results in a new pole formula, based on $\hat{Q}_{1}$ instead of $\hat
{Q}_{6}$. For $G_{8}^{s}/G_{8}<0$, the sign of the interference between the
short-distance and dispersive $\gamma\gamma$ amplitudes in $K_{L}%
\rightarrow\mu^{+}\mu^{-}$ is flipped.

Along the same lines, the pole contribution to $\Delta M_{K}^{LD}$ was shown
to be essentially due to $\hat{Q}_{1}$, pleading again for a better knowledge
of the low-energy constants $x_{i}$, that is to say, of $G_{8}^{s}$%
.\bigskip\newline\textbf{Acknowledgments:} S.T. would like to thank the
organizers of the XLIrst Rencontres de Moriond. J.-M.G. acknowledges support
by the Belgian Federal Office for Scientific, Technical and Cultural Affairs
through IAP P5/27; C.S. is supported by the Schweizerischer Nationalfonds;
S.T. is supported by the DFG grant No.\thinspace NI\thinspace1105/1-1; this
work has also been supported in part by IHP-RTN, EC contract No.\thinspace
HPRN-CT-2002-00311 (EURIDICE).

\end{document}